\documentclass[11pt,a4paper]{article}
\usepackage[T1]{fontenc}
\usepackage[utf8]{inputenc}
\usepackage{authblk}
\usepackage{a4wide}
\usepackage{makeidx}         
\usepackage{graphicx}        
\usepackage{multicol}        
\usepackage[bottom]{footmisc}
\usepackage{hyperref}

\makeindex             

\begin{document}

\title{Longitudinal analysis of gene expression profiles using functional mixed-effects models}
\author[1]{Maurice Berk} 
\author[2]{Cheryl Hemingway}
\author[2]{Michael Levin}
\author[1]{Giovanni Montana\thanks{Email: {\tt g.montana@imperial.ac.uk}}}
\affil[1]{Department of Mathematics, Imperial College London} 
\affil[2]{Division of Medicine, Imperial College London}

\maketitle

\abstract{In many longitudinal microarray studies, the gene expression levels in a random sample are observed repeatedly over time under two or more conditions. The resulting time courses are generally very short, high-dimensional, and may have missing values. Moreover, for every gene, a certain amount of variability in the temporal profiles, among biological replicates, is generally observed. We propose a functional mixed-effects model for estimating the temporal pattern of each gene, which is assumed to be a smooth function. A statistical test based on the distance between the fitted curves is then carried out to detect differential expression. A simulation procedure for assessing the statistical power of our model is also suggested. We evaluate the model performance using both simulations and a real data set investigating the human host response to BCG exposure.}

\section{Introduction}

In a longitudinal microarray experiment, the gene expression levels of a group of biological replicates - for example human patients - are observed repeatedly over time. A typical study might involve two or more biological groups, for instance a control group versus a drug-treated group, with the goal to identify genes whose temporal profiles differ between them. It can be challenging to model these experiments in such a way that accounts for both the within-individual (temporal) and between-individual correlation - failure to do so may lead to poor parameter estimates and ultimately a loss of power and incorrect inference. Further challenges are presented by the small number of time points over which the observations are made, typically fewer than $10$, the high dimensionality of the data with many thousands of genes studied simultaneously, and the presence of noise, with many missing observations.

In order to address these issues we present here a functional data analysis (FDA) approach to microarray time series analysis. In the FDA paradigm we treat observations as noisy realisations of an underlying smooth function of time which is to be estimated. These estimated functions are then treated as the fundamental unit of observation in the subsequent data analysis as in \cite{Ramsay2006}. Similar approaches have been used for the clustering of time series gene expression data without replication \cite{Bar-Joseph2003a,Ma2006} but these cannot be applied to longitudinal designs such as the one described in Section \ref{bcgstudy}. Our approach is much more closely related to, and can be considered a generalisation of, the EDGE model presented by \cite{Storey2005}.

The rest of this paper is organised as follows. Our motivating study is introduced in Section \ref{bcgstudy}. In Section \ref{sec:model} we present our methodology based on functional mixed-effects models. A simulation study is discussed in Section \ref{sec:simul} where we compare our model to EDGE. Section \ref{sec:results} provides a brief summary of our  experimental findings.

\section{A case study: tubercolosis and BCG vaccination} \label{bcgstudy}

Tuberculosis (TB) is the leading cause of death world-wide from a curable infectious disease. In $2006$ it accounted for over $9$ million new patients and over $2$ million deaths; these figures are in spite of effective medication and a vaccine being available since $1921$. This discrepancy is due in part to the HIV-epidemic, government cutbacks, increased immigration from high prevalence areas and the development of multi-drug resistant strains of the disease but ultimately due to our limited understanding of the complex interaction between the host and the pathogen \textit{M. tuberculosis}. In particular, it has been a longstanding observation that the BCG vaccine conveys different levels of protection in different populations \cite{Fine1995}. A total of $17$ controlled trials of the efficacy of BCG vaccination have been carried out and efficacy has varied between $95\%$ and $0\%$; some studies even show a negative effect \cite{Colditz1994}.

\begin{figure}[htbp]
\centering
\includegraphics[width=\textwidth]{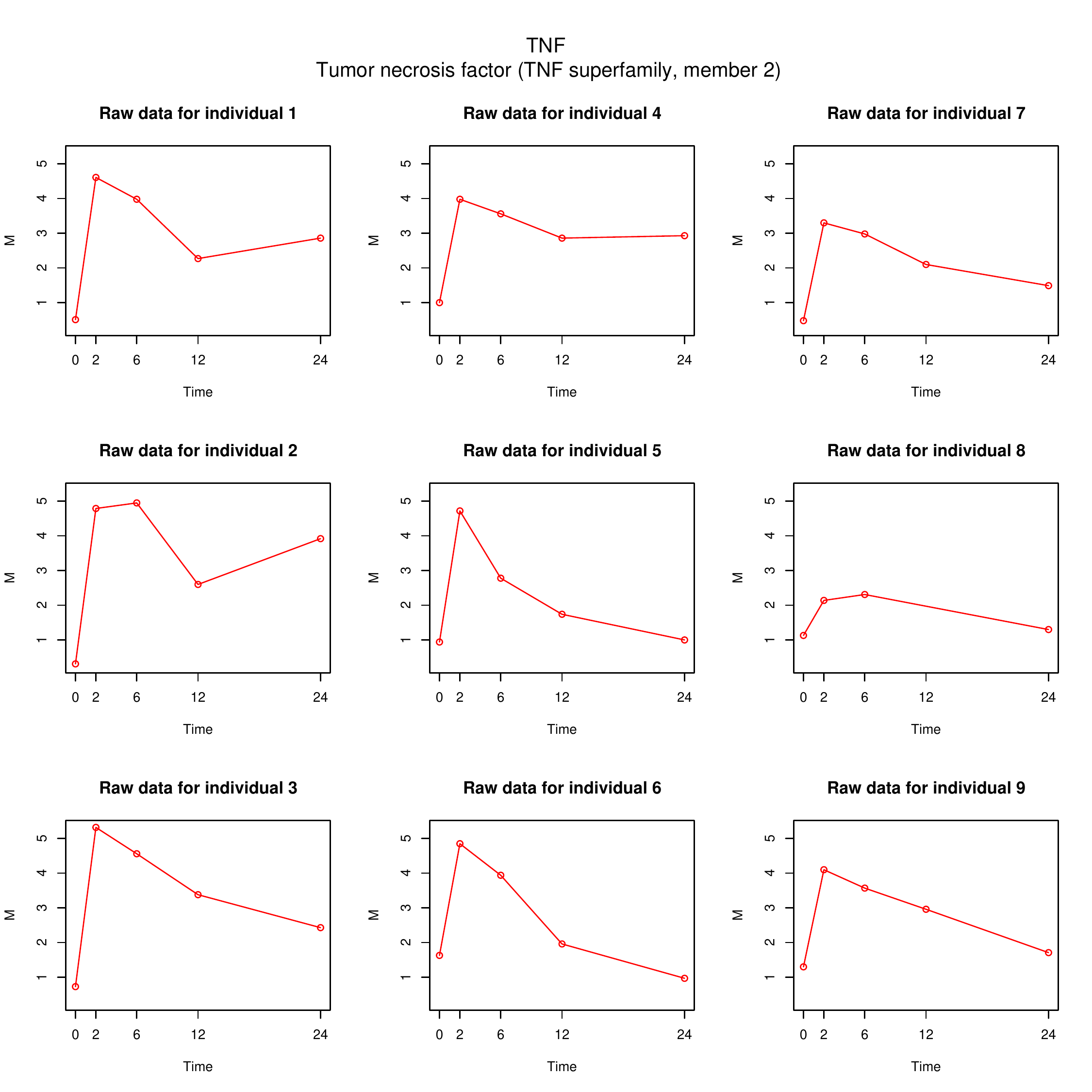}
\caption{An example of $9$ individual gene expression profiles (biological replicates) for the TNF gene. The experimental setup is described in Section \ref{bcgstudy}. Shown here are the raw data, before model fitting. Some of the peculiar features of the data can be observed here: (i) very short temporal profiles, (ii) irregularly spaced design time points, (iii) missing data, and (iv) individual variability.} 
\label{fig:rawdata}
\end{figure}

The purpose of this case study was to characterise the host response to BCG exposure by using microarrays to identify genes which were induced or repressed over time in the presence of BCG. $9$ children with previous exposure to TB but who were then healthy, having completed TB therapy at least $6$ months prior to the study were recruited from the Red Cross Children's Hospital Wellcome TB research database, matched for age and ethnicity. A complete description of the experimental procedures will be reported in a separate publication. In summary, each child contributed a BCG treated and a BCG negative control time series observed at $0, 2, 6, 12$ and $24$ hours after the addition of BCG or, in the case of the controls, $100 \mu$l PBS. A two-colour array platform - the Stanford `lymphochip' - was used. Data prepocessing and quality control were performed using the GenePix4.0 software and in R using BioConductor (\url{www.bioconductor.org}). Figure \ref{fig:rawdata} shows $9$ biological replicates that have been observed for the TNF (tumor necrosis factor) gene, from which three typical features of the longitudinal data under analysis can be noted: (a) all time series are short and exhibit a clear serial correlation structure; (b) a few time points are missing (for instance, individual $8$ has only $4$ time points); (c) there is variability in the gene expression profiles across all individuals. 

\section{Mixed-effects smoothing splines models} \label{sec:model}


Each observation being modelled is denoted by $y(t_{ij})$ and represents the gene expression measure observed on individual $i$ at time $t_{ij}$, where $i=1, 2, \ldots, n_k$,  $j=1,2, \ldots, m_i$, $n_k$ denotes the sample size in group $k$ and $m_{i}$ is the number of observations on individual $i$. In order to properly account for the features observed in Figure \ref{fig:rawdata}, we suggest to model $y(t_{ij})$ non-parametrically: 

\begin{equation} \label{eq:model1}
y(t_{ij}) = \mu(t_{ij}) + v_{i}(t_{ij}) + \epsilon_{ij} 
\end{equation}

\noindent The model postulates that the observed gene expression measure $y(t_{ij})$ can be explained by the additive effect of three components: a mean response $\mu(\cdot)$, which is assumed to be a smooth, non-random curve defined over the time range of interest; an individual-specific deviation from that mean curve, $v_{i}(\cdot)$, which is assumed to be a smooth and random curve observed over the same time range; and an error term $\epsilon_{ij}$ which accounts for the variability not explained by the first two terms. Formally, we treat each $v_{i}(t_{ij})$, for $i=1, 2, \ldots, n_k$, as independent and identically distributed realisations of an underlying stochastic process; specifically, we assume that $v_{i}(t_{ij})$ is a Gaussian process with zero-mean and covariance function $\gamma(s,t)$, that is $v_{i}(t_{ij}) \sim \textrm{GP}(0, \gamma)$. The errors terms $\epsilon_{ij}$ are assumed to be independent and normally distributed with zero mean and covariance matrix $\vec{R}_i$. We do not assume that all individual have been observed at the same design time points, and all the distinct design time points are denoted by $(\tau_1, \tau_2, \ldots, \tau_m)$.


We suggest to represent the curves using cubic smoothing splines; see, for instance, \cite{Green1994}. The key idea of smoothing splines consists in making full use of all the design time points and then fitting the model by adding a smoothness or roughness constraint; by controlling the size of this constraint, we are then able to avoid curves that appear too wiggly. A natural way of measuring the roughness of a curve is by means of its integrated squared second derivative, assuming that the curve is twice-differentiable. We call $\vec \eta =(\eta(\tau_1), \dots, \eta(\tau_m))^T$ the vector containing the values of the mean curve estimated at all design time points and, analogously, the individual-specific deviations from the mean curve, for individual $i$, are collected in $\vec v_i = (v_i(\tau_1), \ldots, v_i(\tau_m))^T$. The mean and individual curves featuring in model (\ref{eq:model1}) can be written as, respectively, 
$\mu(t_{ij}) = \vec x_{ij}^T \vec \eta$ and $v_{i}(t_{ij}) = \vec x_{ij}^T \vec v_{i}$, with $i=1,2,\ldots,n$, and $\vec x_{ij} =(x_{ij1}, \ldots, x_{ijm})^T$, with $x_{ijr}=1$ if $t_{ij}=\tau_r, ~ r=1,\dots,m$ and $x_{ijr}=0$ otherwise. The fact that the individual curves are assumed to be Gaussian processes is captured by assuming that the individual deviations are random and follow a zero-centred Gaussian distribution with covariance $\vec D$, where $\vec D(r,s) =\gamma(\tau_s, \tau_r)$, $r,s=1,\ldots,m$. Finally, in matrix form, model (\ref{eq:model1}) can then be rewritten as

\begin{eqnarray}
\label{eq:model2}
& \vec y_i = \vec X_i \vec \eta + \vec X_i \vec v_i + \vec \epsilon_i &\\
\nonumber & \vec v_i \sim \textrm{N}(\vec 0, \vec D) \qquad \vec \epsilon_i \sim \textrm{N}(\vec 0, \vec R_i) &
\end{eqnarray}

For simplicity, we assume that $\vec R_i = \sigma^2 \vec I$. 
In this form, we obtain a linear mixed-effects model \cite{Laird1982}. Clearly, the model accounts for the fact that, for a given gene, the individual repeated measurements are correlated. Specifically, under the assumptions above, we have that $\textrm{cov}(\vec y_i) = \vec X_i \vec D \vec X_i^T + \vec R_i$.

\subsection{Statistical inference}

A common approach to estimating the unknown parameters of a linear mixed-effects model is by maximum likelihood (ML) estimation. In our model (\ref{eq:model2}), the twice negative logarithm of the (unconstrained) generalised log-likelihood for is given by
$$
\sum_{i=1}^{n_k} \left\{ (\vec y_i - \vec X_i \vec \eta - \vec X_i \vec v_i ) ^T \vec R_i^{-1} ( \vec y_i - \vec X_i \vec \eta - \vec X_i \vec v_i ) + \log \mid \vec D \mid + \vec v_i^T \vec D^{-1}  \vec v_i + \log \mid \vec R_i \mid \right\}. 
$$

The ML estimators of the mean curve $\mu(\cdot)$ and each individual curve $v_i(\cdot)$ can be obtained by minimising a penalised version of this log-likelihood obtained by adding a term $\lambda \vec \eta^T \vec G \vec \eta$ and a term $\lambda_v \sum_{i=1}^{n_k} \left\{ \vec v_i ^T \vec G \vec v_i \right\}$, which impose a penalty on the roughness of the mean and individual curves, respectively. The matrix $\vec G$ is the roughness matrix that quantifies the smoothness of the curve \cite{Green1994} whilst the two scalars $\lambda_v$ and $\lambda$ are smoothing parameters controlling the size of the penalties. In principle, $n_k$ distinct individual smoothing parameters can be introduced in the model but such a choice would incur a great computational cost during model fitting and selection. For this reason, we assume that, for each given gene, all individual curves share the same degree of smoothness and we use only one smoothing parameter $\lambda_{v}$.

After a rearrangement on the terms featuring in the penalised log-likelihood, the model can be re-written in terms of the regularised covariance matrices, $\vec D_v  = (\vec D^{-1} + \lambda_v \vec G)^{-1}$ and $\vec V_i  = \vec X_i \vec D_v \vec X_i^T + \vec R_i$. When both these variance components are known, the ML estimators $\hat{\vec \eta}$ and $\hat{\vec v}_i$, for $i=1,2,\dots,n_{k}$, can be derived in closed-form as the minimisers of the penalised generalised log-likelihood. However, the variance components are generally unknown. All parameters can be estimated iteratively using an EM algorithm, which begins with some initial guesses of the variance components. The smoothing parameters $\lambda$ and $\lambda_{v}$ are found as those values, in the two-dimensional space $(\Lambda \times \Lambda_v)$, that optimise the \textit{corrected} AIC, which includes a small sample size adjustment. The search for optimal smoothing values $\hat \lambda$ and $\hat \lambda_{v}$ is performed using a downhill simplex optimisation algorithm \cite{Nelder1965}.

The objective of our analysis is to compare the estimated mean curves observed under the two experimental groups and assess the null hypothesis that the curves are the same. After fitting model (\ref{eq:model2}) to the data, independently for each group and each gene, we obtain the estimated mean curves $\hat \mu^{(1)}(t)$ and $\hat \mu^{(2)}(t)$. One way of measuring the dissimilarity between these two curves consists in computing the $L_2$ distance between them, which can be evaluated using the smoothed curves $\hat \mu^{(1)}(t)$ and $\hat \mu^{(2)}(t)$, thus yielding the observed distance $\hat D$. We use this dissimilarity measure as a test statistic. Since the null distribution of this statistic is not available in closed form, we resort to a non-parametric bootstrap approach in order to approximate it. 

\begin{figure}[htbp]
\centering
\includegraphics[width=\textwidth]{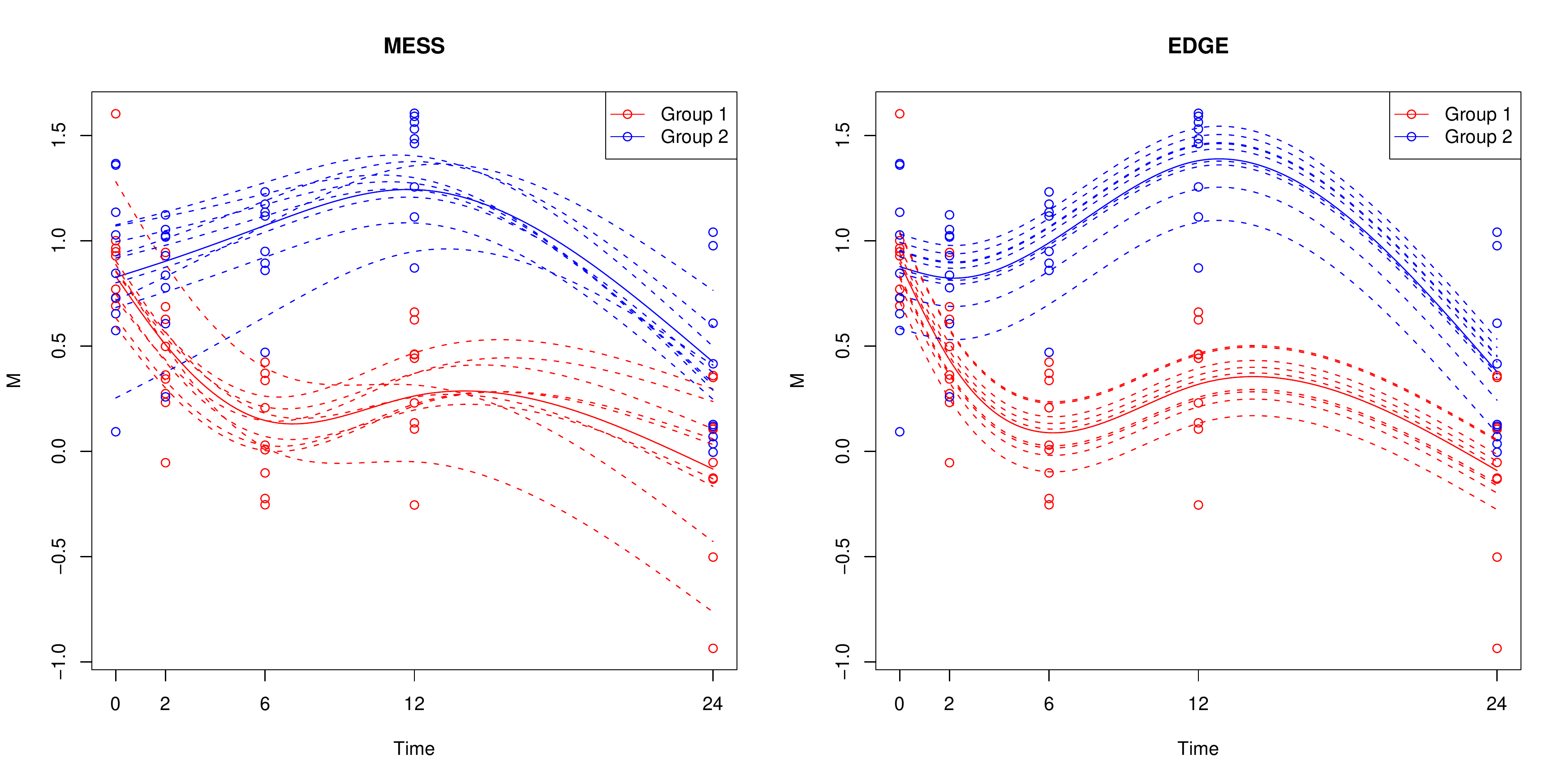}
\caption{An example of simulated longitudinal data and fitted curves using both MESS and EDGE. The thick solid lines correspond to the fitted means for each group. The dotted lines are the fitted individual curves for group 1 and the dashed lines are the fitted individual curves for group 2.}
\label{fig:simulated}
\end{figure}

\section{Performance assessment using simulated longitudinal data} \label{sec:simul}

In order to assess the performance of the proposed MESS model we compared it using a simulation study to the EDGE model developed by \cite{Storey2005}. While the EDGE model takes the same form as (\ref{eq:model1}), their parameterisation differs from ours in that the mean function $\mu(\cdot)$ is represented using B-splines and the individual curves $v_i(\cdot)$ are constrained to be a scalar shift. In the case of the mean curve, the B-spline representation requires specification of both the number and location of the knots which, unlike smoothing splines, offers discontinuous control over the degree of smoothing. Furthermore \cite{Storey2005} represent each gene using the same number of basis functions which, if taken to be too small, implies a poor fit to those genes with the most complex temporal profiles. Conversely, if the number of basis functions is sufficient to model these complex genes there is a danger that some genes will be overfit. In the case of the individual curves $v_i(\cdot)$, it should be clear that scalar shifts would be unable to fully capture the individual variability we observe in the raw data given in Figure \ref{fig:rawdata}. This problem is compounded by the fact that \cite{Storey2005} propose an F-type test statistic for inference which makes use of the model residuals.

To determine the practical impact of these features we have set up a simulation procedure that generates individual curves that look similar to the real experimental data. Our procedure is based on a mixed-effects model with the fixed- and random-effects parameterized using B-splines, where the observations on individual $i$ belonging to group $j$ are given as
\begin{eqnarray}
\label{eqn:1}
& \vec{y}_{i} = \vec{X}_{i}\vec{\beta}_{j} + \vec{Z}_{i}\vec{b}_{ij} + \vec{\epsilon}_{ij} &\\
\nonumber & \vec{b}_{ij} \sim \textrm{MVN}(\vec{0},\vec{D}_{j}) \qquad \vec{\epsilon}_{ij} \sim  \textrm{MVN}(\vec{0},\sigma_{j}\vec{I}_{n_{i} \times n_{i}}) &
\end{eqnarray}
where $i=1,\ldots,n$ and $j=1,2$. For simplicity, we use the same basis for the fixed- and random-effects so that $\vec{X}_{i} = \vec{Z}_{i}$. The parameters that need to be controlled in this setting therefore consist of the variance components $\sigma_{1}, \sigma_{2}, \vec{D}_{1}, \vec{D}_{2}$, the B-spline parameters for the group means $\vec{\beta}_{1}$, $\vec{\beta}_{2}$, and the B-spline basis $\vec{X}_{i}$ which is determined by the number and locations of the knots, $K$. Given the simple patterns often observed in real data sets, we place a single knot at the center of the time course. Wherever possible, we tune the variance components based on summary statistics computed from data produced in real studies such as the experiment described in Section \ref{bcgstudy}.  We further parameterise the covariance matrices as
\begin{eqnarray}
& \vec{D} = \tau \left[ \begin{array}{cccc}
\rho^{0} & \rho^{1} & \rho^{2} & \cdots \\
\rho^{1} & \rho^{0} & \rho^{1} & \cdots \\
\rho^{2} & \rho^{1} & \rho^{0} & \ddots \\
\vdots & \ddots & \ddots & \ddots
\end{array}
\right] &
\end{eqnarray}
and introduce the notation $\vec{D}(\tau,\rho)$ for specifying these parameters. In this way we can vary the complexity of the individual curves by varying the parameter $\rho$ and control the amount of variation between individuals by varying $\tau$. When $\rho=1$, the individual `curves' are scalar shifts, as in the EDGE model. As $\rho$ tends to $0$, $\vec{D}$ tends to $\tau\vec{I}$, where the B-spline parameters $\vec{b}_{i}$ are independent.

We begin simulating a given gene by first randomly generating the B-spline co-efficients for the mean curve of group 1, $\vec{\beta}_{1}$, and for each individual belonging to this group, $\vec{b}_{i1}$, according to the following probability distributions $\vec{\beta}_{1} \sim \textrm{MVN}(\vec{0}, \vec{D}_{\beta})$ and $\vec{b}_{i1} \sim \textrm{MVN}(\vec{0},\vec{D}_{b_{1}})$,  with covariance matrices given by $\vec{D}_{\beta} = \vec{D}(0.25,0.6)$ and $\vec{D}_{b_{1}} = \vec{D}(\tau_{b_{1}},0.6)$, where $\tau_{b_{1}}  \sim U(0.1,0.2)$. As in (\ref{eqn:1}), the error term is normally distributed with variance $\sigma_{1}$. We set this variance component to be log-normally distributed with mean $-2$ and variance $0.35$, values estimated from the real data.

Each simulated gene is differentially expressed with probability $0.1$. If a gene is not differentially expresed then observations are generated for group 2 using exactly the same parameters as for group 1, i.e. $\vec{\beta}_{1}=\vec{\beta}_{2}, \vec{D}_{b_{1}}=\vec{D}_{b_{2}}, \sigma_{1}=\sigma_{2}$. On the other hand, if a gene is differentially expressed, then we randomly generate a vector $\vec{\beta}_{\delta}$ representing the difference in B-spline parameters for the group mean curves, distributed as
$\vec{\beta}_{\delta} \sim \textrm{MVN}(\vec{0}, \vec{D}_{\vec{\beta}_{\delta}})$ and $\vec{D}_{\vec{\beta}_{\delta}} = \vec{D}(0.25,0.9)$,
with $\beta_{\delta1} = 0$. We then normalise the vector $\vec{\beta}_{\delta}$ so that its $L_{2}$-norm is $1$ before setting $\vec{\beta}_{2} = \vec{\beta}_{1} + \vec{\beta}_{\delta}$. By setting $\beta_{\delta1} = 0$ we ensure that both mean curves began the time course at the same value, which we have observed in the real data and would similarly be the case if the data had been $t=0$ transformed. Setting $\rho=0.9$ for $\vec{D}_{\vec{\beta}_{\delta}}$ limits the difference between the curves in terms of complexity which, again, we observe in the real data where frequently the mean curves are simply vertically shifted versions of each other. Normalising the vector $\vec{\beta}_{\delta}$ enables us to control exactly how large an effect size we are trying to detect by multiplying the vector by a scaling factor. 

Finally, we generate the individual curves for group 2 for a differentially expressed gene as before:
$\vec{b}_{i2} \sim \textrm{MVN}(\vec{0},\vec{D}_{b_{2}})$ and $\vec{D}_{b_{2}} = \vec{D}(\tau_{b_{2}},0.6)$, where $\tau_{b_{2}} \sim U(0.1,0.2)$. The key point to note is that $\tau_{b_{2}} \ne \tau_{b_{1}}$. By doing so, a differentially expressed gene varies both in terms of the mean curve and the degree of individual variation. Similarly, $\sigma_{2}$ is distributed identically to yet independently of $\sigma_{1}$ so that the noise of the two groups is also different.

\begin{figure}[htbp]
\centering
\includegraphics[scale=0.35]{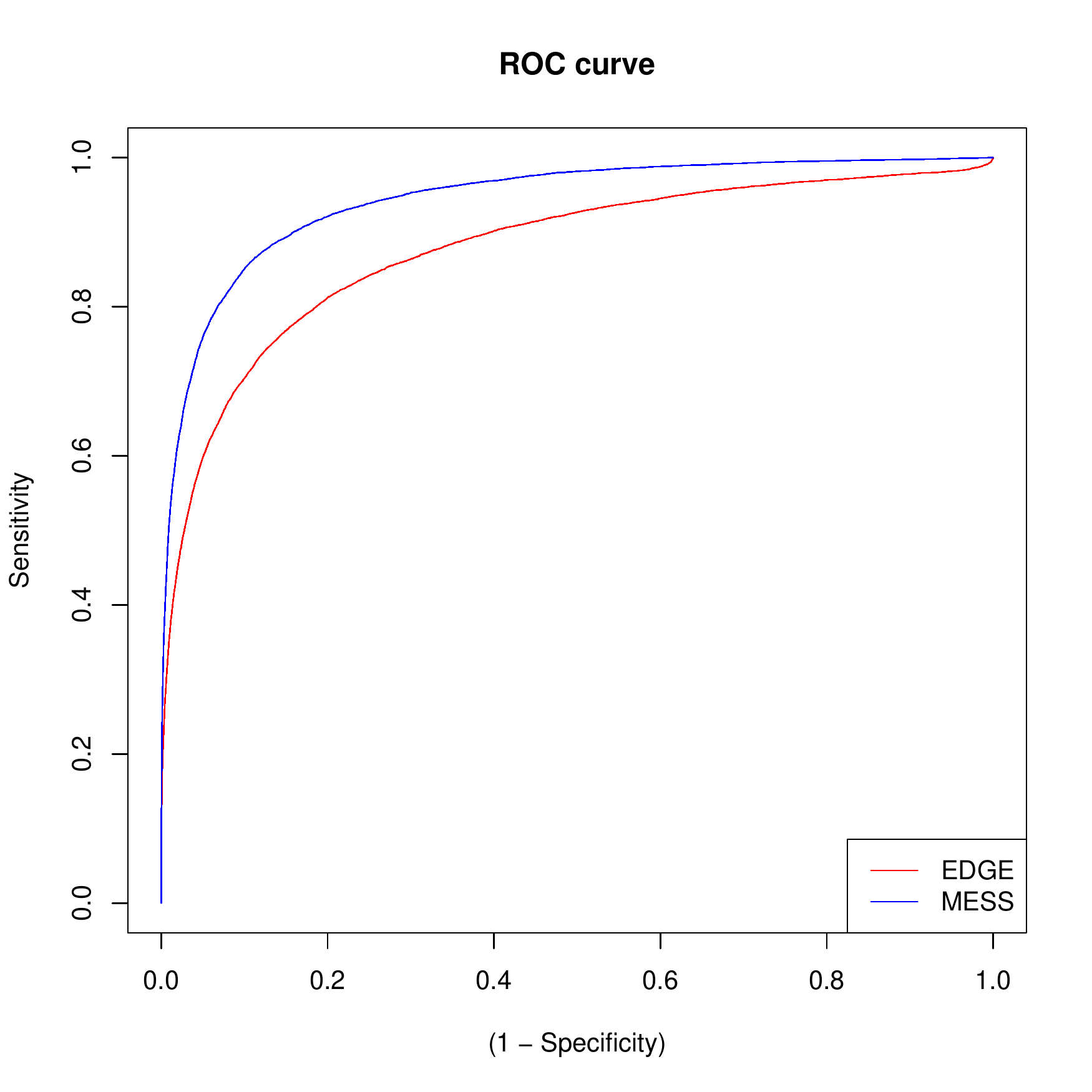}
\caption{ROC curve comparing the performance between the EDGE and MESS models. Results are based on $100,000$ simulated genes as described in Section \ref{sec:simul}.} 
\label{fig:roc}
\end{figure}

\begin{figure}[htbp]
\includegraphics[width=\textwidth]{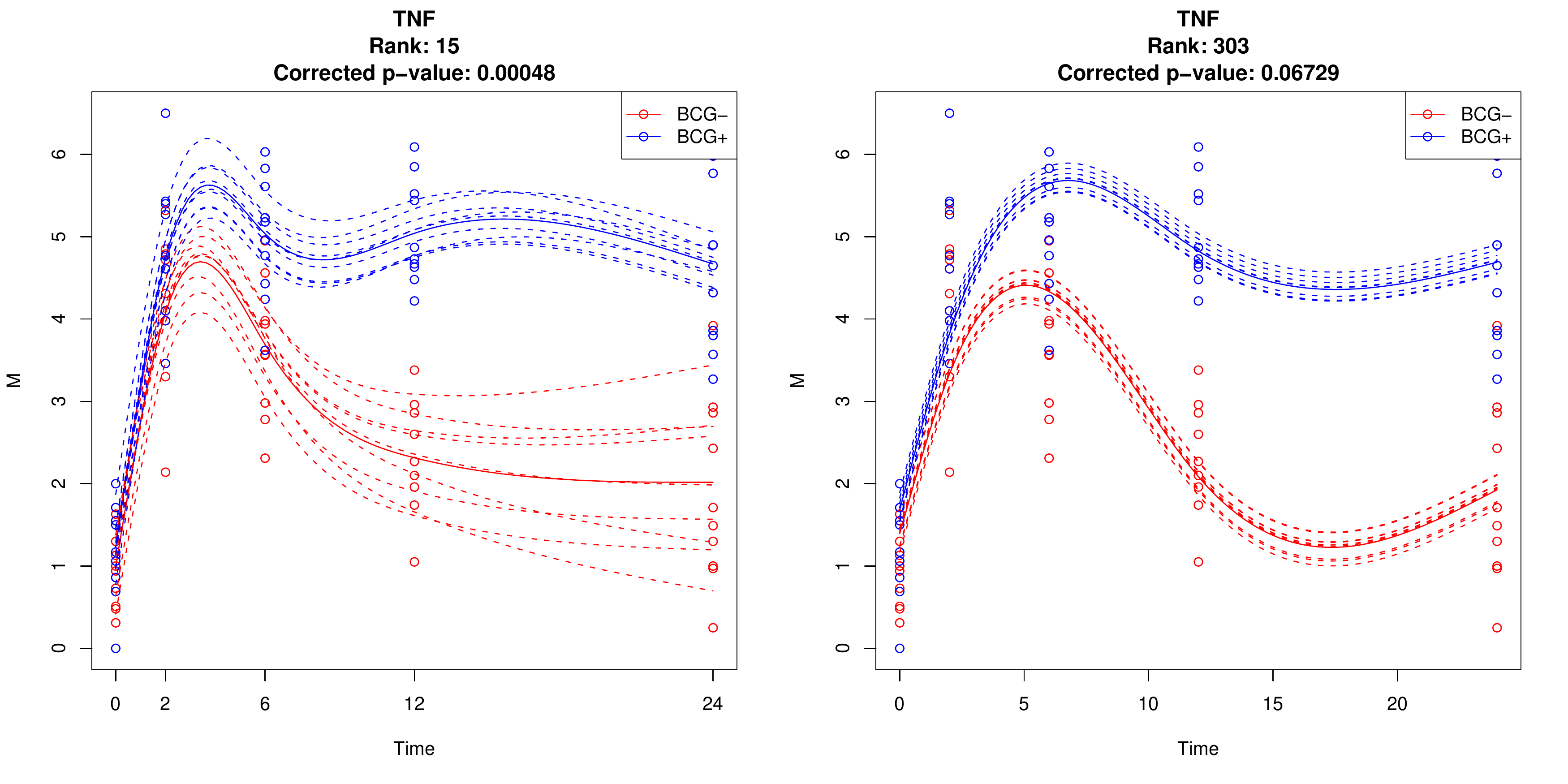}
\caption{BCG case study: a top scoring genes according to MESS (left), but not to EDGE (right). TNF has been strongly implicated in TB infection \cite{Flynn1995} and we would expect it to be ranked as highly significant. EDGE's low ranking can be partly explained by poor model selection failing to accurately capture the gene expression dynamics, and the inadequacy of scalar shifts to fully explain the variation between individuals. }
\label{fig:result}
\end{figure}

Using this simulation framework with the parameters laid out as above, we generated a data set consisting of $100,000$ simulated genes observed with 9 individuals per group with 5 timepoints at $0, 2, 6, 12$ and $24$ hours, following the same pattern of observations as the case study. $10\%$ of genes were differentially expressed. We then used both the MESS and EDGE models to fit the data and identify differentially expressed genes. Figure \ref{fig:simulated} shows an example of a simulated gene with fitted mean and individual curves for both MESS and EDGE. In this instance EDGE's B-spline parameterisation seems sufficient for representing the mean curves but the scalar shifts do not model the data as closely as MESS does. Compare this simulated gene to a real example from the experimental data shown in Figure \ref{fig:result}. This is the fit to the gene TNF, for which the raw data for the control group was given in Figure \ref{fig:rawdata}. We can see here that EDGE has selected too few knots to adequately capture the rapid spike in gene expression levels at 2 hours and that again the MESS model with individual curves provides a much closer fit to the data. Figure \ref{fig:roc} gives the ROC curve for the simulation study based on $100,000$ simulated genes. At a fixed specificity of $90\%$, the corresponding power for MESS is $85.1\%$ compared to $70.4\%$ for EDGE.

\section{Experimental results}\label{sec:results}

We fit the MESS model to the BCG case study data and generated 100 bootstrap samples giving 3.2 million null genes from which to calculate empirical p-values based on the $L_{2}$ distance as a test statistic. After correcting these p-values for multiple testing, 359 probes were identified as being significantly differentially expressed, corresponding to 276 unique genes. We provide here a brief summary of the results, leaving the full biological interpretation to a dedicated forthcoming publication.

The top ten differentially expressed genes were found to be CCL20, PTGS2, SFRP1, IL1A, INHBA, FABP4, TNF, CXCL3, CCL19 and DHRS9. Many of these genes have previously been identified as being implicated in TB infection. For instance, CCL20 was found to be upregulated in human macrophages infected with \textit{M.tuberculosis} \cite{Ragno2001} and \textit{in vivo} in patients suffering from pulmonary TB \cite{Lee2008}, while TNF-$\alpha$ has had a long association with the dieseae \cite{Flynn1995}. In total, using the GeneCards online database (\url{www.genecards.org}), 58 of the 276 significant genes were found to have existing citations in the literature corresponding to \textit{M.tuberculosis} infection or the BCG vaccine. Those which were upregulated mainly consisted of chemokines and cytokines such as CCL1, CCL2, CCL7, CCL18, CCL20, CXCL1, CXCL2, CXCL3, CXCL9, CXCL10, TNF, CSF2, CSF3, IFNG, IL1A, IL1B, IL6 and IL8 while the downregulated genes were exemplified by transmembrame receptors CD86, CD163, TLR1, TLR4 and IL8RB. The large number of differentially expressed genes that we would have expected to identify lends credence to those genes without citations and whose role in the host response to BCG is currently unknown.

\section{Conclusions}

In this work we have presented a non-parametric mixed-effects model based on smoothing splines for the analysis of longitudinal gene expression profiles. Experimental results based on both simulated and real data demonstrate that the use of both a flexible model that incorporates individual curves and an appropriate test-statistic yields higher statistical power than existing functional data analysis approaches.


\section*{acknowledgement}
The authors thank the Wellcome Trust for their financial support

\bibliographystyle{abbrv}
\bibliography{biblio}

\end{document}